\documentclass[conference]{IEEEtran}
\usepackage{cite}
\usepackage{amsmath,amssymb,amsfonts}
\usepackage{algorithmic}
\usepackage{graphicx}
\usepackage{textcomp}
\usepackage{xcolor}
\usepackage{pifont}
\usepackage{subcaption}
\usepackage{array}    
\usepackage{hyperref}
\def\BibTeX{{\rm B\kern-.05em{\sc i\kern-.025em b}\kern-.08em
    T\kern-.1667em\lower.7ex\hbox{E}\kern-.125emX}}
\begin{document}

\title{Two-Stage Camera Calibration Method for Multi-Camera Systems Using Scene Geometry\\
\thanks{Identify applicable funding agency here. If none, delete this.}
}

\author{\IEEEauthorblockN{Aleksandr Abramov}
\IEEEauthorblockA{\textit{Mallsense} \\
Dubai, United Arab Emirates \\
a.abramov@mallsense.ae}
\IEEEauthorblockA{Co-founder and CTO}
}

\maketitle

\begin{abstract}
Calibration of multi-camera systems is a key task for accurate object tracking. However, it remains a challenging problem in real-world conditions, where traditional methods are not applicable due to the lack of accurate floor plans, physical access to place calibration patterns, or synchronized video streams. This paper presents a novel two-stage calibration method that overcomes these limitations. In the first stage, partial calibration of individual cameras is performed based on an operator's annotation of natural geometric primitives (parallel, perpendicular, and vertical lines, or line segments of equal length). This allows estimating key parameters (roll, pitch, focal length) and projecting the camera's Effective Field of View (EFOV) onto the horizontal plane in a base 3D coordinate system. In the second stage, precise system calibration is achieved through interactive manipulation of the projected EFOV polygons. The operator adjusts their position, scale, and rotation to align them with the floor plan or, in its absence, using virtual calibration elements projected onto all cameras in the system. This determines the remaining extrinsic parameters (camera position and yaw). Calibration requires only a static image from each camera, eliminating the need for physical access or synchronized video. The method is implemented as a practical web service. Comparative analysis and demonstration videos confirm the method's applicability, accuracy, and flexibility, enabling the deployment of precise multi-camera tracking systems in scenarios previously considered infeasible.
\end{abstract}


\section{Introduction}
Video surveillance systems for object tracking and monitoring are used in diverse environments – both indoor and outdoor – utilizing static or moving cameras. The range of tasks addressed includes monitoring critical infrastructure, analyzing personnel behavior in industrial settings, and tracking visitors in retail spaces.

The development of this field is driven by two key factors: the availability of a mature algorithmic base and the continuous reduction of computational costs due to advancements in GPU technology. This makes video tracking cost-effective for a wide range of applications where its implementation was previously economically unviable.

The requirements for tracking vary depending on the task: from simple detection of an object's presence to detailed analysis of its dynamics and individual parameters. In many scenarios, comprehensive processing of data from multiple cameras is necessary, which requires building an end-to-end multi-camera tracking system. A key element of such a system is the precise coordination of data between cameras.

For tasks requiring detailed behavioral analysis, such as assessing visitor movements in a shopping mall, accurate real-time estimates of each object's position and motion parameters are critically important. This enables the generation of detailed data for subsequent spatial efficiency analysis.

The quality of multi-camera tracking directly depends on the accuracy of camera calibration. Correct calibration allows transitioning from processing data in individual camera coordinate systems to analysis in a unified 3D coordinate system of the location. This enables track refinement when an object is simultaneously observed by multiple cameras, seamless handover of objects between cameras, and the ability to predict trajectories in areas between non-overlapping fields of view. The practical effectiveness of a tracking system built on precise calibration is demonstrated in video surveillance systems that implement person detection and tracking with projection of their positions onto a floor plan \cite{demo_tracking_plan}, tracking of the body's centerline \cite{demo_centerline}, and seamless merging of tracks from adjacent cameras \cite{demo_merging_tracks}.

\section{Challenges in Multi-Camera Calibration}

Despite the significant number of existing camera calibration methods \cite{salvi2002comparative, zhang2008wide}, their applicability heavily depends on the specific use case. Various approaches, based on different mathematical models and algorithms, impose different requirements on input data and provide varying levels of accuracy. Consequently, the correct choice of calibration method becomes a determining factor for the successful deployment of a system in real-world conditions across various application scenarios. Several main approaches to multi-camera system calibration can be distinguished:

\begin{itemize}
    \item Pattern-based calibration
    \item Map-based calibration
    \item Tracking-based calibration
    \item Camera network calibration
    \item Deep learning models
\end{itemize}

\begin{table*}[htbp]
\centering
\caption{Comparison of multi-camera calibration approaches}
\label{tab:calibration_comparison}
\begin{tabular}{|l|c|c|c|c|c|}
\hline
\textbf{Requirements and Results} & \textbf{Pattern-based} & \textbf{Map-based} & \textbf{Tracking-based} & \textbf{Network} & \textbf{Deep Learning} \\
\hline
Need for: placement patterns under cameras & $\bullet$ &  &  &  &  \\
\hline
Need for: an accurate floorplan &  & $\bullet$ &  &  &  \\
\hline
Need for: video objects tracking &  &  & $\bullet$ & $\circ$ &  \\
\hline
Need for: marking control elements on the image &  & $\bullet$ &  & $\circ$ &  \\
\hline
Need for: marking control elements on the floorplan &  & $\bullet$ &  &  &  \\
\hline
Need for: visual control of calibration quality & $\circ$ & $\bullet$ & $\circ$ & $\circ$ & $\circ$ \\
\hline
Result: approximate mutual camera correspondence & $\bullet$ & $\bullet$ & $\bullet$ & $\bullet$ & $\circ$ \\
\hline
Result: plane match of each camera and a floorplan & $\bullet$ & $\bullet$ & $\circ$ &  & $\circ$ \\
\hline
Result: full accurate match image plane and 3D coordinates & $\bullet$ & $\circ$ &  &  & $\circ$ \\
\hline
\end{tabular}
\end{table*}

These approaches impose different requirements on data and operator actions:

\begin{itemize}
    \item Need for placing calibration objects within the field of view
    \item Processing of video streams and object tracking
    \item Availability of an accurate floor plan
    \item Marking control markers on the plan and image
    \item Visual control of calibration quality
\end{itemize}

The approaches also differ in their final result:

\begin{itemize}
    \item An approximate correspondence between camera coverage areas is achieved
    \item The coordinate correspondence between the camera image plane and the location plan is achieved
    \item The complete coordinate correspondence between the camera image coordinate system and the unified 3D coordinate system of the location is achieved
\end{itemize}

A comparative analysis of the described calibration approaches is presented in Table \ref{tab:calibration_comparison}, which summarizes their requirements and achieved results. The table uses the following notation: $\circ$ indicates partial applicability of a method to the specific requirement, while $\bullet$ denotes full applicability.

The \textbf{pattern-based calibration} approach involves placing a specific template in front of the camera, such as a flat or three-dimensional object with known parameters (dimensions, pattern geometry, QR code, etc.). Subsequently, based on the known information about the pattern object, the required parameters for each camera's calibration are determined. A key feature of this approach is the requirement for physical access to the camera and the ability to capture images with the pattern placed in front of it \cite{zhang2000flexible, wu2015lens}. When using high-quality patterns, this approach can achieve high accuracy in calibration.

The \textbf{map-based calibration} approach requires that you have a plan of the location where the camera system operates. This is followed by either automatic detection or manual marking of pairs of control points (or other elements), where one element of the pair lies on the camera image and the other element is specified on the location plan. Based on these pairs of associated elements, camera parameters are estimated and the desired coordinate correspondence is built. A key feature of this approach is the requirement for a high-quality location plan and the presence of distinct points and lines in the image that can be correlated with their counterparts on the plan \cite{zhang2008wide}. If the requirements for the quality of the location plan are met, this method can also achieve high accuracy in the calibration.

The \textbf{tracking-based calibration} approach involves processing synchronous video footage from multiple cameras, where moving objects are detected and tracked. As objects move, the sizes of their projections on the cameras change, providing a basis for estimating both the relative positions of the cameras and the individual optical and positional parameters of each camera. A characteristic of this approach is the need for synchronous video processing, as well as non-trivial algorithms for object detection and tracking \cite{zhang2008wide}. As a result, given a sufficient amount of calibration data, the method allows one to achieve an approximate coordinate correspondence.

The \textbf{camera network calibration} approach involves building a model of the cameras' relative positions considering adjacent overlapping areas. In the overlapping zones of neighboring fields of view, control elements visible from both cameras are marked \cite{zhang2008wide}. This provides a basis for understanding the approximate relative positioning of the cameras. A feature of this method is the necessity of marking control elements on the camera images. Applying this method results in building an approximate correspondence for the boundary areas of the cameras.

The \textbf{deep learning model} approach is relatively new and involves performing calibration using a pre-trained neural network model. A key feature of this approach is that it requires only a single image from the camera and has potential applicability across a wide range of scenarios \cite{liao2023deep}. However, currently, this approach does not lead in terms of accuracy and leaves open the question of precise result correction when calibration inaccuracies are detected.

This work considers a scenario for deploying a multi-camera system where obtaining the data required for applying existing calibration methods is difficult or impossible. Such scenarios include retail, industrial, and other locations where accurate floor plans are unavailable, or it is impossible to obtain synchronous video recordings for tracking. Under these conditions, traditional approaches, as well as their combinations, prove to be inapplicable. At the same time, the ability to quickly and accurately deploy a video tracking system in such challenging conditions is often a critical factor determining the project's profitability and overall success.

The application of existing calibration methods in real-world conditions often turns out to be impossible due to a number of fundamental limitations. Inaccuracies in location plans arise from frequent environmental changes: rearrangement of temporary walls and retail equipment, modification of production lines, etc. As a result, the plan used for calibration no longer corresponds to the actual environment observed by the cameras. In other cases, a plan may be entirely absent or unavailable due to commercial or industrial security reasons. Physical access for placing calibration patterns is also often impossible - either due to strict security protocols at the facility or the high cost of such work. This problem is particularly critical when analyzing archival footage from locations that no longer exist. These same limitations make it impossible to record synchronous video from multiple cameras, which is necessary for methods based on tracking moving objects.

\section{Proposed Calibration Method}

The proposed method belongs to the class of approaches based on the analysis of visible scene geometry \cite{echigo1989camera, echigo2008perspective, nakano2020camera}. Unlike methods requiring special calibration patterns or accurate floor plans, this approach utilizes natural geometric elements observed within the camera's field of view (perspective lines, architectural features, interior elements). These visual cues contain sufficient information for estimating camera parameters, which is particularly valuable in conditions where the application of traditional methods is impossible or limited.

The method implements a two-stage calibration procedure. In the first stage, partial calibration of individual cameras is performed based on an operator's annotation of specific geometric primitives. Although not all camera parameters are determined at this stage, the obtained approximation is sufficient for projecting the fields of view into a unified coordinate space. In the second stage, precise mutual calibration of the camera system is achieved through interactive manipulation of the projected fields of view using virtual calibration elements. Reference primitives (e.g., coordinate markers of a specific shape) are created in the space, which the operator can move while simultaneously observing their projection onto the images from all cameras in the system. This allows for visual identification and correction of calibration discrepancies, ensuring complete control over the camera alignment process.

The theoretical foundations of the method, the algorithms used, its advantages and limitations are discussed in detail below. A description of the implemented web service is also provided, which offers tools for performing calibration, visualizing spatial geometry, and interactively correcting camera parameters using virtual calibration elements.

\begin{figure}[htbp]
\centering
\includegraphics[width=1.0\linewidth]{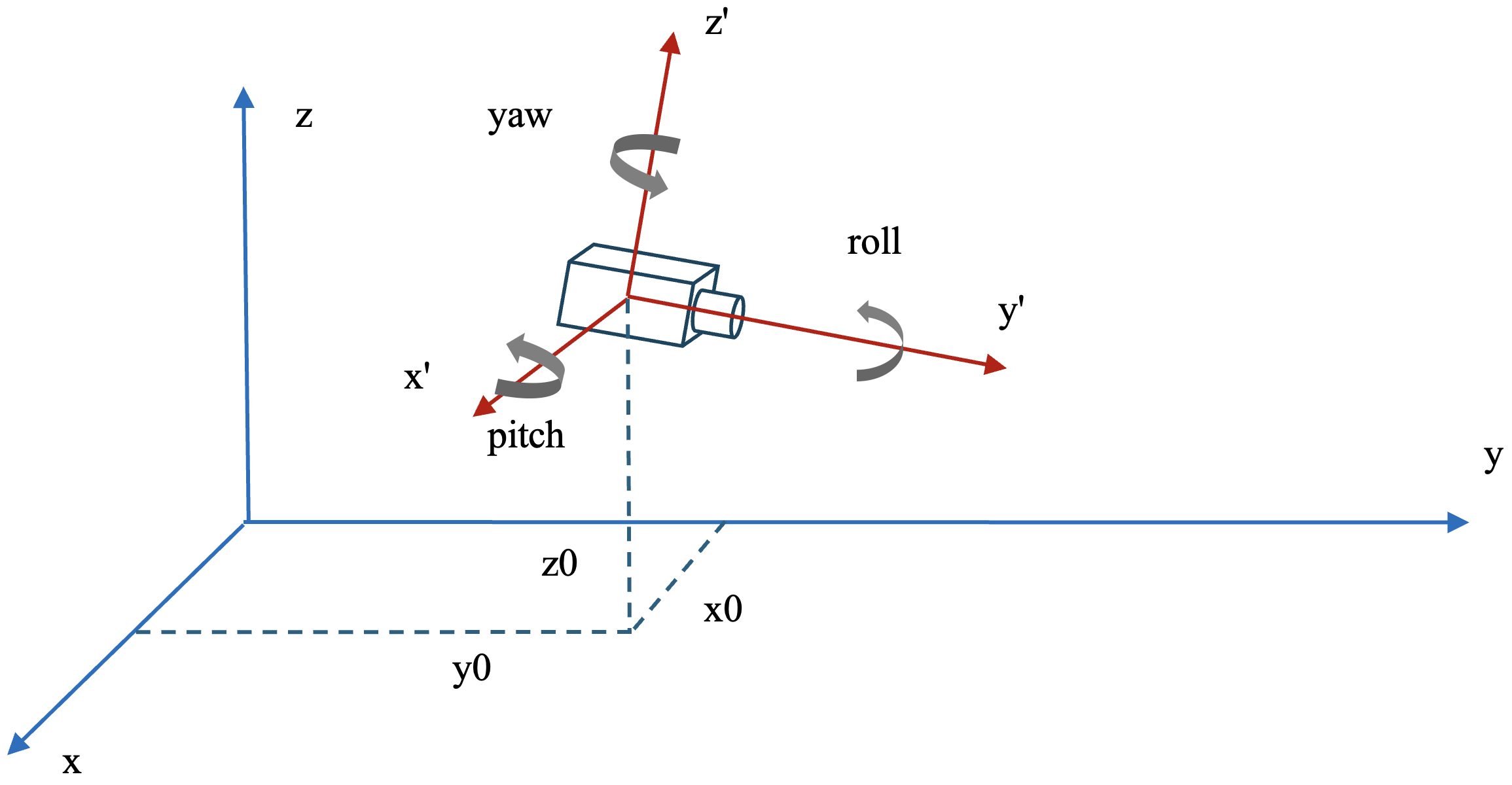}
\caption{Relationship between the base CS (blue) and the camera-associated CS (red). The coordinate systems are linked via the displacement vector (x0, y0, z0) and a sequence of three rotations: yaw, pitch, roll.}
\label{fig:coordinate_systems}
\end{figure}

\begin{figure}[htbp]
\centering
\includegraphics[width=1.0\linewidth]{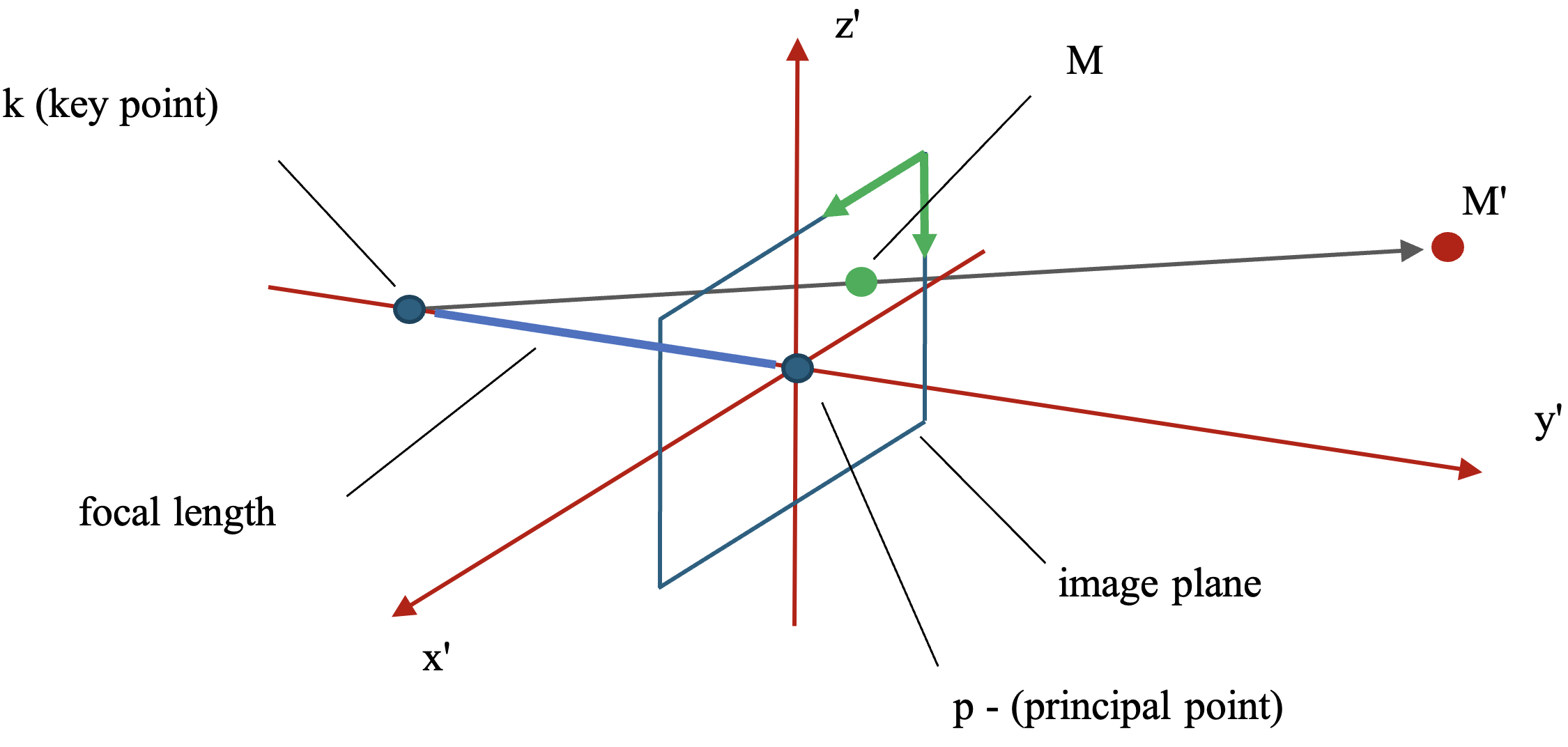}
\caption{Schematic of the projection of point M', defined in the camera-associated CS (x', y', z') (red), onto the image plane M, defined in the pixel CS of the camera image (green). Point M' is obtained at the intersection of the projection ray from point k to point M with the image plane.}
\label{fig:projection_schematic}
\end{figure}

The method is based on the well-known Pinhole Camera Model, commonly used in the field of camera calibration. This model serves to link the image coordinate system (hereinafter referred to as CS) \(\left(x^{pix}, y^{pix}\right)\), and the location-associated 3D coordinate system (\(x\), \(y\), \(z\)), hereafter called the base system. Linking these coordinate systems enables a coordinate transition from the image to the 3D CS and vice versa, allowing one to obtain the image point for any point in real space (Fig.~\ref{fig:coordinate_systems} and Fig.~\ref{fig:projection_schematic}).

Thus, through sequential coordinate transformations, this model provides the capability to find the projection of any spatial point, defined in the base CS, onto the image, and conversely, to project any point on the image into the base CS, if the height of the projection plane is known. Let us denote the functions of these transformations as follows:

\begin{flalign*}
& X_{pix} = f^{3d \to pix}(X_{3d}, x0, y0, z0, roll, pitch, yaw, focal) & \\
& X_{3d} = f^{pix \to 3d}(X_{pix}, h, x0, y0, z0, roll, pitch, yaw, focal) &
\end{flalign*}

Where:
\begin{itemize}
    \item $X_{pix}$ - point in the two-dimensional pixel CS of the image;
    \item $X_{3d}$ - point in the base three-dimensional CS;
    \item $f^{3d \to pix}$ - coordinate transformation function from the base 3D CS to the pixel CS of the camera image;
    \item $f^{pix \to 3d}$ - coordinate transformation function from the pixel CS of the camera image to the base 3D CS;
    \item $x0, y0, z0, roll, pitch, yaw, focal$ - parameters of the pinhole camera model: three components of the camera position, three camera rotation angles, focal length;
\end{itemize}

Since any point in the pixel CS of the camera image corresponds not to a single point in the base 3D CS, but to a set of points on the ray $\overline{kp}$ (Fig.~\ref{fig:projection_schematic}), for an unambiguous correspondence, specifying the height $h$ of the projection plane for the transformation function $f^{pix \to 3d}$ is required. This height defines the horizontal plane at height $h$, at the intersection with the ray $\overline{kp}$ of which lies the desired projection point. If the ray $\overline{kp}$ is not parallel to this plane, then such a projection point can be found.

The key idea of the proposed calibration method is to use existing visible perspective geometry lines for camera calibration. Since the application area of the method is considered for static indoor cameras, the presence of geometric elements with such informational potential is widespread. The calibration process is divided into two stages, the sequential execution of which leads to the complete calibration of the camera system. In the first stage, the operator performing the system setup is prompted to outline on the camera image certain sets of geometric lines that obey specific spatial arrangement rules (note that adherence to these rules is not required in the image itself). Additionally, the operator is also prompted to outline on the camera image the polygon of the camera's effective field of view (EFOV) in the floor plane. Then, based on the annotated lines, part of the camera model parameters is computed, and using these partially computed parameters, the projection of the field of view polygon into the base CS is performed. In the second stage, the projections of the effective field of view polygons for each camera are visualized for the operator on the location plan, and they are prompted to perform three types of actions: moving, rotating, and scaling these polygons. Based on the displacement, rotation, and scale resulting from these actions, the remaining part of the camera model parameters is determined. The stages of the method are described in detail below.

\textbf{Stage 1}. The operator is prompted to outline lines on each camera's image that obey specific geometric rules. At this stage, two different options for annotating geometry lines on the camera image are assumed.

\subsubsection{First Annotation Option}
In the first option, the operator must outline a series of lines (Fig.~\ref{fig:annotation_option1}):
\begin{itemize}
    \item A set of lines parallel to each other in space, lying in the floor plane
    \item A pair of lines perpendicular to each other in space, lying in the floor plane
    \item A set of vertical lines in space
\end{itemize}

\subsubsection{Second Annotation Option}
\begin{figure*}[htbp]
\centering
\includegraphics[width=0.5\linewidth]{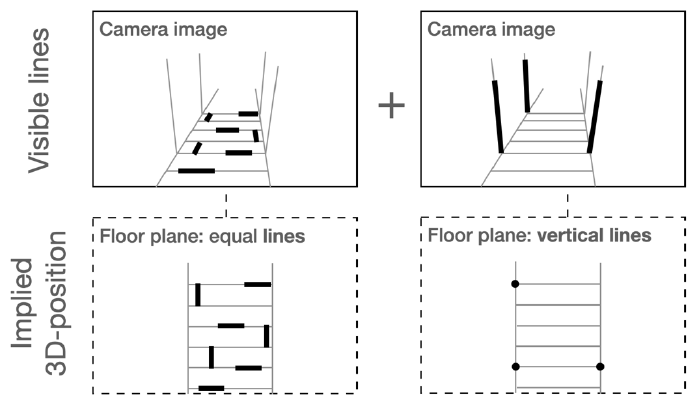}
\caption{Second option for annotating visible perspective geometry lines on the camera image.}
\label{fig:annotation_option2}
\end{figure*}

In the second option, the operator must outline a series of lines (Fig.~\ref{fig:annotation_option2}):
\begin{itemize}
    \item A set of segments equal in length to each other in space, located in the floor plane
    \item A set of vertical lines in space
\end{itemize}

\begin{figure*}[htbp]
\centering

\begin{minipage}{0.4\textwidth}
\centering
\includegraphics[width=\linewidth]{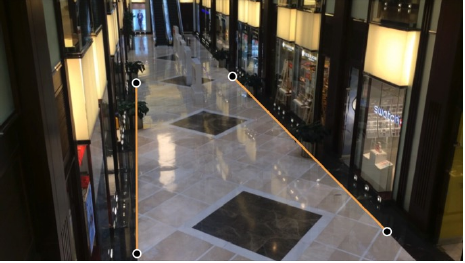}
\subcaption{Implied position: parallel lines}
\end{minipage}
\hspace{1.0cm}
\begin{minipage}{0.4\textwidth}
\centering
\includegraphics[width=\linewidth]{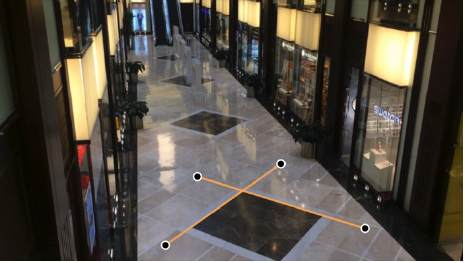}
\subcaption{Implied position: perpendicular lines}
\end{minipage}

\vspace{0.2cm}

\begin{minipage}{0.4\textwidth}
\centering
\includegraphics[width=\linewidth]{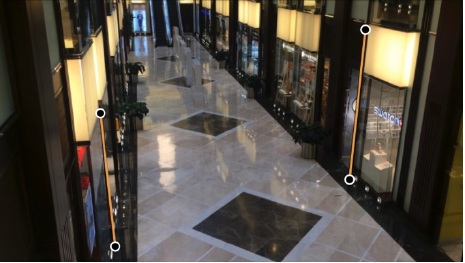}
\subcaption{Implied position: vertical lines}
\end{minipage}
\hspace{1.0cm}
\begin{minipage}{0.4\textwidth}
\centering
\includegraphics[width=\linewidth]{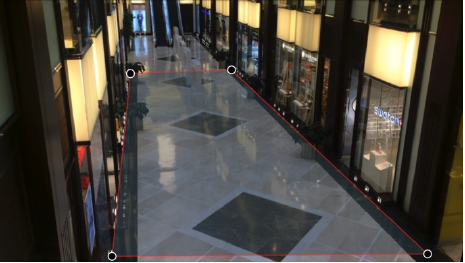}
\subcaption{EFOV polygon}
\end{minipage}

\caption{Examples of geometric line annotation according to the first option.}
\label{fig:annotation_examples}
\end{figure*}

Adherence to the geometric rules in space is ensured by the operator's skill, and in most cases, for indoor cameras, such elements of visible geometry are easily found. Compliance with these rules is required precisely in the base CS, meaning the operator, based on their own understanding of the field of view's geometry, assumes that the marked lines obey certain rules, even though in the image CS they do not obey these same rules. Furthermore, the implied arrangement of the line sets is often well distinguishable and obvious from visual cues. Vertical lines are well distinguishable at the boundaries of walls, corners, doorways, and other interior elements. Parallel lines and lines perpendicular to each other on the floor plane are well distinguishable along the boundaries of walls and floors, by the relative position of columns, and by paving elements. Segments of equal length, in the floor plane, can also be distinguished by paving elements, by the distance between columns, and by other visible elements. Notably, the start and end points, when marking a group of vertical lines, as well as a group of parallel lines and a pair of perpendicular lines, do not matter; that is, it is not important where these lines start and end, only their directions are important. Fig.~\ref{fig:annotation_examples} shows an example of annotating a real camera image according to the first option.

\begin{figure*}[htbp]
\centering
\includegraphics[width=0.7\linewidth]{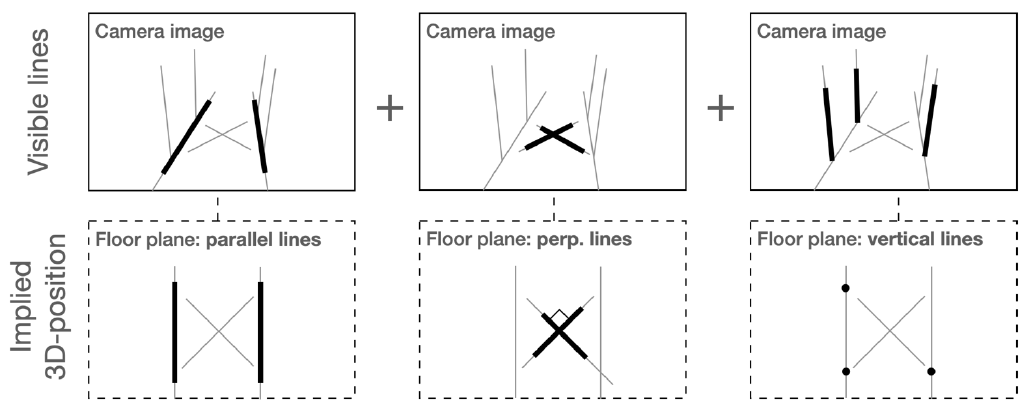}
\caption{First option for annotating visible perspective geometry lines on the camera image.}
\label{fig:annotation_option1}
\end{figure*}

After the line sets have been outlined according to one of the two options (Fig.~\ref{fig:annotation_option1} or Fig.~\ref{fig:annotation_option2}, respectively), the computation of part of the camera model parameters is performed. At this stage, the parameters $roll$, $pitch$, $focal$ are estimated.

\begin{figure}[htbp]
\centering
\begin{minipage}[t]{0.5\textwidth}
\centering
\includegraphics[width=1.0\linewidth]{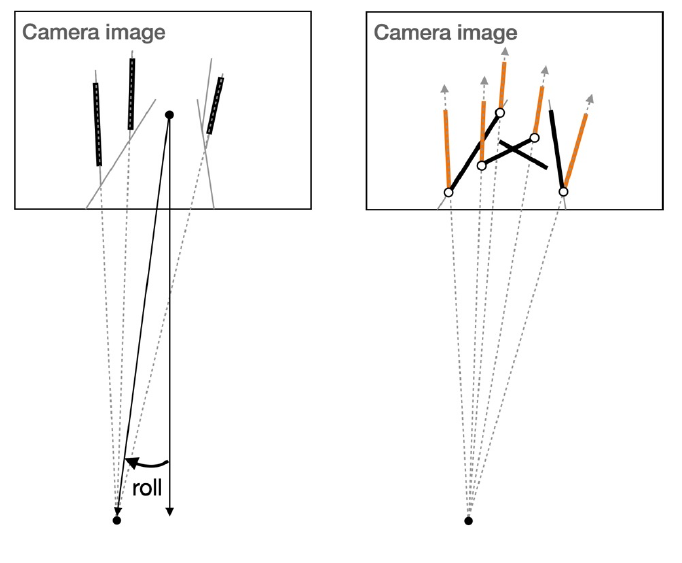}
\end{minipage}
\caption{Finding the vanishing point on the image of annotated lines, vertical in space, and computing the $roll$ angle (left). Using the found vanishing point to construct other vertical lines in space, based on the coordinates of any point lying in the horizontal plane (right).}
\label{fig:vanishing_point_roll}
\end{figure}

First, the $roll$ angle -- the camera rotation angle around the optical axis -- is determined. Since both annotation options involve outlining several lines on the image that are vertical in space (and thus parallel to each other), these lines will have a vanishing point on the image. By finding this vanishing point, the desired $roll$ rotation angle can subsequently be computed (Fig.~\ref{fig:vanishing_point_roll}, left). This vanishing point can also be used to construct other vertical lines in space, based on the coordinates of any point lying in the horizontal plane (Fig.~\ref{fig:vanishing_point_roll}, right).

\vspace{0.4cm}

\vspace{0.4cm}

After the $\widehat{roll}$ rotation angle is determined, the $pitch$ rotation angle and the focal length $focal$ can be determined by solving an optimization problem using numerical optimization methods.

\begin{flalign*}
& \widehat{pitch}, \widehat{focal} = & \\
& \underset{pitch, focal}{\text{argmin}} \, C(x0, y0, z0, \widehat{roll}, pitch, yaw, focal) &
\end{flalign*}

Since the parameters $x0$, $y0$, $z0$, $yaw$ are not defined at this point, they are set to default values -- camera position with zero offset at a height of three meters.

\[
\widehat{pitch}, \widehat{focal} = \underset{pitch, focal}{\text{argmin}} \, C(0, 0, 3, \widehat{roll}, pitch, 0, focal)
\]

Next, consider the optimization criterion $C(x0, y0, z0, roll, pitch, yaw, focal)$ or $C(*ps)$ for brevity, which is chosen such that its minimum corresponds to the combination of transformation parameters that leads to the fulfillment of the geometric conditions in the base CS space when projecting the lines annotated by the operator on the camera image into this CS. Let's consider an example of constructing the criterion, assuming the operator annotated lines according to option 1: two lines parallel in space $pl1$ and $pl2$, two lines perpendicular in space $pp1$ and $pp2$, and two lines vertical in space $vl1$ and $vl2$. Then the criterion $C$ takes the form:

\begin{flalign*}
& C(*ps) = V(vl1, vl2, *ps) & \\
& \quad + \left|1 - \left|\text{dot}\left(f^{pix \to 3d}(pl1, 0, *ps), f^{pix \to 3d}(pl2, 0, *ps)\right)\right|\right| & \\
& \quad + \left|\text{dot}\left(f^{pix \to 3d}(pp1, 0, *ps), f^{pix \to 3d}(pp2, 0, *ps)\right)\right| &
\end{flalign*}

Where:
\begin{itemize}
    \item $*ps$ - parameters defining the camera model $[x0, y0, z0, roll, pitch, yaw, focal]$
    \item $V(vl1, vl2, *ps)$ -- a function that takes the value 0 if the lines $vl1$, $vl2$, projected into the base CS are parallel, and a value $> 0$ if parallelism is not maintained.
\end{itemize}

Thus, the criterion $C(x0, y0, z0, roll, pitch, yaw, focal)$ takes a value equal to zero if the geometric conditions in space are met: $pl1$ and $pl2$ turn out to be parallel as a result of projection into the base CS, $pp1$ and $pp2$ turn out to be perpendicular, $vl1$ and $vl2$ also turn out to be parallel to each other and perpendicular to the horizontal plane. Then solving the optimization problem:

\[
\widehat{pitch}, \widehat{focal} = \underset{pitch, focal}{\text{argmin}} \, C(0, 0, 3, \widehat{roll}, pitch, 0, focal)
\]

yields optimal estimates for $\widehat{pitch}$, $\widehat{focal}$. As a result, it is possible to build a partially defined (in the sense that some parameters are set by default, and some are computed) coordinate correspondence:

$X_{3d} = f^{pix \to 3d}(X_{pix}, h, x0, y0, z0, roll, pitch, yaw, focal)$ in the form:

\[
X_{3d} = f^{pix \to 3d}(X_{pix}, h, 0, 0, 3, \widehat{roll}, \widehat{pitch}, 0, \widehat{focal})
\]

\begin{figure}[htbp]
\centering
\includegraphics[width=0.97\linewidth]{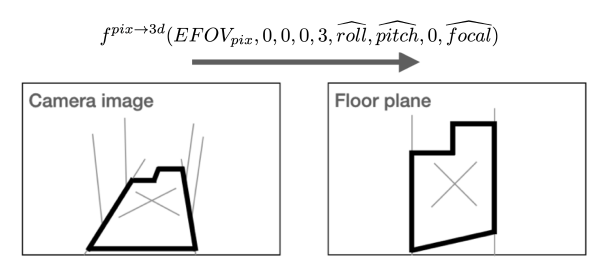}
\caption{Projection of the effective field of view polygon, outlined on the camera image, into the base CS using partial calibration.}
\label{fig:efov_projection}
\end{figure}

Such a partially defined coordinate correspondence turns out to be sufficient to project the effective field of view polygon (EFOV -- Effective Field Of View) outlined by the operator in the floor plane into the base CS in a form undistorted by perspective transformation (Fig.~\ref{fig:efov_projection}).

\[
EFOV_{3d} = f^{pix \to 3d}(EFOV_{pix}, 0, 0, 0, 3, \widehat{roll}, \widehat{pitch}, 0, \widehat{focal})
\]

Such a polygon projection $EFOV_{3d}$ will have correct proportions, accurate up to the displacement $(x0, y0)$, mounting height $z0$, and the camera's horizontal rotation angle $yaw$. These parameters, missing for full calibration, are to be estimated in the second calibration stage.

In the case of the second annotation option, where segments of equal length are marked instead of parallel lines in the floor plane and perpendicular lines, the criterion is formed according to a similar principle, ensuring the estimation of optimal camera model parameters leading to the fulfillment of geometric conditions as a result of projecting the annotated elements into the base CS.

\begin{figure}[htbp]
\centering
\includegraphics[width=0.94\linewidth]{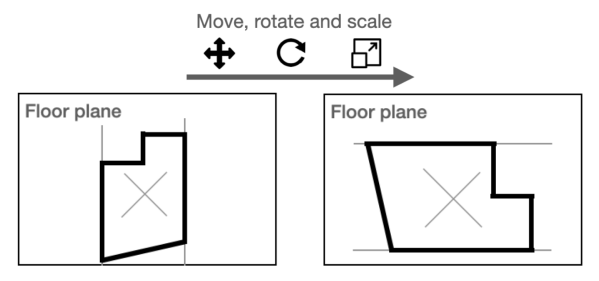}
\caption{Moving, scaling, and rotating the polygon, leading to the determination of the missing camera model parameters.}
\label{fig:polygon_manipulation}
\end{figure}

\textbf{Stage 2}. In the second calibration stage, the operator begins working with the projected EFOV polygons on the horizontal plane of the base CS. The shapes of these polygon projections are not distorted by perspective; however, further determination of the camera position and orientation parameters is required: $x0$, $y0$, $z0$, $yaw$. The operator is prompted to perform actions on the polygons: move, scale, and rotate them, achieving the desired camera position in the base CS (Fig.~\ref{fig:polygon_manipulation}).

\begin{figure*}[htbp]
\centering
\includegraphics[width=0.8\linewidth]{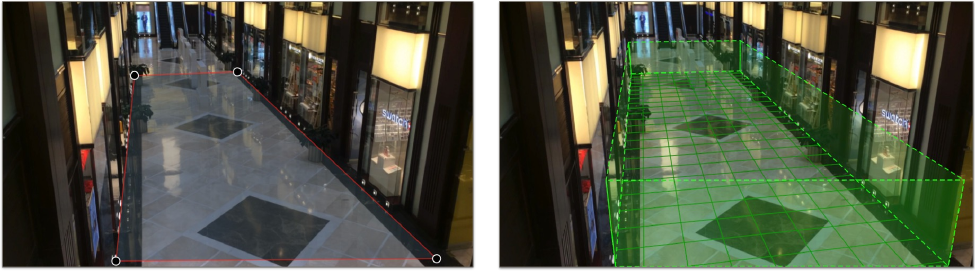}
\caption{The camera's effective field of view polygon EFOV (left), and the 3D reconstruction of the two-meter high prism based on it (right).}
\label{fig:3d_fov_reconstruction}
\end{figure*}

The parameters are determined according to the following principle. The displacement $x0$, $y0$ is determined as a result of moving the polygon relative to the initial projection point. Since the polygon at this stage is defined in the base CS, the performed displacement will correspond to the camera position coordinates. The default mounting height was set to 3 meters. Accordingly, the scaling factor applied to the polygon during the operator's scaling action is determined, and the mounting height is multiplied by this same factor. Thus, if the operator scaled the polygon by a factor of 1.5, the camera mounting height $z0$ is determined as $3 \times 1.5 = 4.5$ meters. Rotating the polygon determines the camera's horizontal rotation angle $yaw$ according to a similar principle. As a result of fixing the polygon in its new position, the missing coordinate correspondence parameters $\widehat{x0}$, $\widehat{y0}$, $\widehat{z0}$, $\widehat{yaw}$ are determined. Thus, the complete direct and inverse coordinate correspondence is formed:

\[
X_{pix} = f^{3d \to pix}(X_{3d}, \widehat{x0}, \widehat{y0}, \widehat{z0}, \widehat{roll}, \widehat{pitch}, \widehat{yaw}, \widehat{focal})
\]
\[
X_{3d} = f^{pix \to 3d}(X_{pix}, h, \widehat{x0}, \widehat{y0}, \widehat{z0}, \widehat{roll}, \widehat{pitch}, \widehat{yaw}, \widehat{focal})
\]

As a result of estimating the camera model parameters, it is possible to perform back-projection using $f^{3d \to pix}$ of the camera's 3D field of view as a prism, whose base is defined by the EFOV polygon, and the top face is obtained by raising the base in the base CS to a height of two meters and back-projecting it into the camera image CS (Fig.~\ref{fig:3d_fov_reconstruction}).

\begin{figure*}[htbp]
\centering
\includegraphics[width=0.8\linewidth]{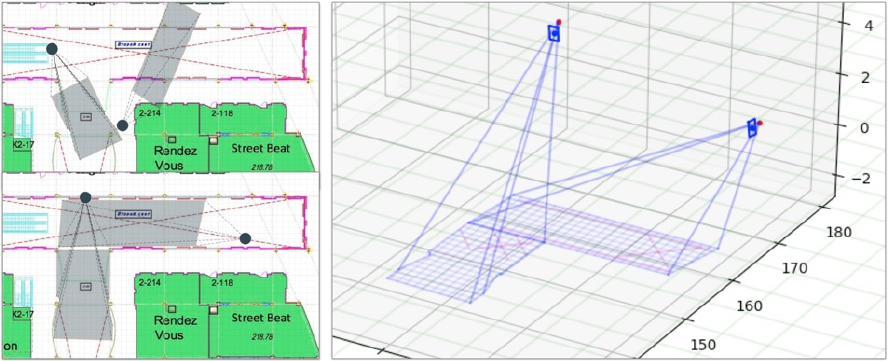}
\caption{Example of aligning EFOV polygons with their real positions on the plan, and building the coordinate correspondence for the camera system.}
\label{fig:efov_alignment}
\end{figure*}

\begin{figure*}[htbp]
\centering
\includegraphics[width=0.9\linewidth]{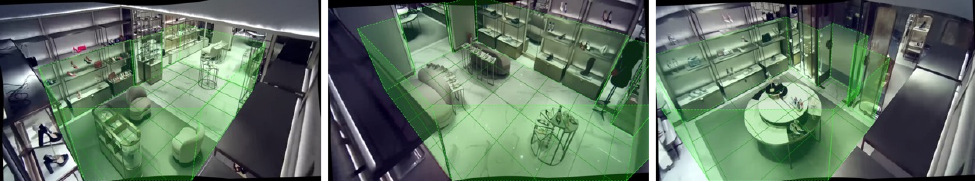}
\hfill
\includegraphics[width=0.9\linewidth]{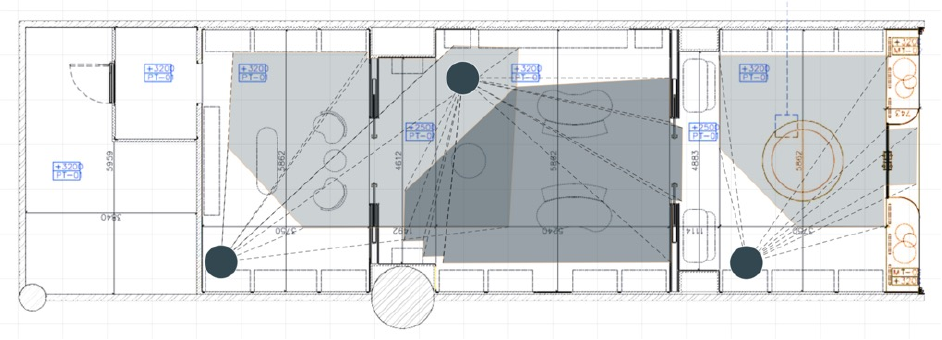}
\hfill
\includegraphics[width=0.9\linewidth]{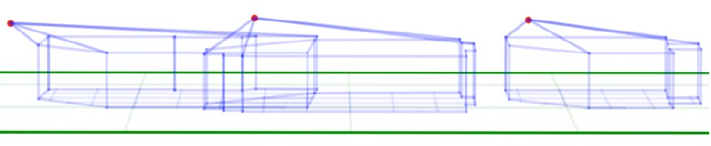}
\caption{Example of aligning three cameras on a unified location plan, and building the coordinate correspondence for the camera system.}
\label{fig:multi_camera_alignment}
\end{figure*}

If a location plan is available, and the camera's position can be determined from its field of view, the EFOV projection polygon on the horizontal plane of the base CS can be moved, stretched, and rotated by the operator so that the polygon's position matches its correct position on the plan (Fig.~\ref{fig:efov_alignment} and Fig.~\ref{fig:multi_camera_alignment}).

If a location plan is not available, the proposed calibration method also provides a way to achieve the result. In this case, moving the field of view polygon to the desired position is impossible, and then a virtual calibration element can be used to calibrate the camera system. This refers to a certain object modeled in the base CS and projected onto all cameras in the system. Using such virtual calibration elements, it is possible to achieve full calibration of the camera system in the complete absence of any plan. An example of applying the calibration method under such conditions for a three-camera system is provided in the demonstration video \cite{demo_no_plan}. Furthermore, even if a location plan is available, such virtual elements can be used for visual control and achieving high calibration accuracy. The application of this approach is demonstrated by the example of calibrating a three-camera system for a cow farm \cite{demo_cow_farm}.

Since the proposed method at the first stage involves annotating straight perspective geometry lines, this means that radial image distortion must be eliminated before applying this method. In the examples above, distortion was preliminarily eliminated by initially annotating polylines over curved lines, which should become straight after distortion removal, and subsequently solving an optimization problem to select such radial distortion model coefficients that would ensure the straightening of these polylines. However, preliminary distortion removal can also be performed by other means, for example \cite{prescott2002line, hartley2019parameter}. In this case, tangential distortion parameters can be included in the camera model and optimized simultaneously with other model parameters.

To enable the application of this calibration method in real-world conditions, an application providing camera system calibration functionality has been developed. This service is implemented as a web application consisting of a frontend part implemented in JavaScript (React) and a backend part implemented in Python (Django, Numpy, Scipy, OpenCV). This application provides an interactive user interface and allows performing the described stages of camera calibration. The service provides visualization of the 3D geometric reconstruction of the camera's effective field of view, coordinate grid, and calibration elements. Such visualization allows interactively performing the necessary actions for adjusting calibration parameters and explicitly controlling the quality of the result. The view of the application's user interface, as well as a demonstration of its operation, are presented in the demo videos mentioned above \cite{demo_no_plan,demo_cow_farm}.

A Russian patent \cite{patent_ru} was obtained for the proposed calibration method, and national applications have been filed in several countries via a PCT application. A US patent \cite{patent_us} was also obtained.

\begin{table*}[htbp]
\centering
\caption{Comparison of multi-camera calibration approaches}
\label{tab:calibration_comparison_final}
\begin{tabular}{|l|c|c|c|c|c|}
\hline
\textbf{Requirements and Results} & \textbf{Pattern-based} & \textbf{Map-based} & \textbf{Tracking-based} & \textbf{...} & \textbf{Proposed method} \\
\hline
Need for: placement patterns under cameras & $\bullet$ &  &  &  &  \\
\hline
Need for: an accurate floorplan &  & $\bullet$ &  &  & $\circ$ \\
\hline
Need for: video objects tracking &  &  & $\bullet$ &  &  \\
\hline
Need for: marking control elements on the image &  & $\bullet$ &  &  & $\bullet$ \\
\hline
Need for: marking control elements on the floorplan &  & $\bullet$ &  &  &  \\
\hline
Need for: visual control of calibration quality & $\circ$ & $\bullet$ & $\circ$ &  & $\bullet$ \\
\hline
Result: approximate mutual camera correspondence & $\bullet$ & $\bullet$ & $\bullet$ &  & $\bullet$ \\
\hline
Result: plane match of each camera and a floorplan & $\bullet$ & $\bullet$ & $\circ$ &  & $\bullet$ \\
\hline
Result: full accurate match image plane and 3D coordinates & $\bullet$ & $\circ$ &  &  & $\bullet$ \\
\hline
\end{tabular}
\end{table*}

\section{Conclusions}

This work has addressed the problem of calibrating multi-camera video surveillance systems in
challenging practical conditions where the application of traditional methods is impossible or limited. As a
solution, an innovative two-stage method was proposed, which effectively overcomes the key limitations
of existing approaches. A comparative Table \ref{tab:calibration_comparison_final} of the methods shows the key advantages and scientific-technical achievements of the method:

\begin{enumerate}
    \item \textbf{Versatility of Application.} The method does not depend on the availability of an accurate floor plan, making it applicable in dynamically changing environments (retail spaces, production facilities), as well as at sites where plans are unavailable or classified. Calibration is possible both with and without a floor plan.
    
    \item \textbf{Operation in Limited Access Conditions.} The approach eliminates the need for physical placement of calibration patterns within the camera's field of view. This is critically important for facilities with strict security protocols or when working with archival footage.
    
    \item \textbf{Minimal Data Requirements.} Calibration requires only a single static image from each camera. The method does not require synchronized video recording, complex object tracking, or object recognition, significantly lowering the entry barrier and computational costs at the deployment stage.
    
    \item \textbf{Intuitive Control and High Accuracy.} The two-stage procedure, combining automated parameter estimation based on visible geometry and subsequent interactive correction by an operator, enables high accuracy. The use of virtual calibration elements provides transparent and clear control over the process, allowing compensation for potential errors.
    
    \item \textbf{Practical Implementation.} The developed web service confirms the method's viability, providing a complete toolkit for calibration, distortion correction, visualization, and result control through a convenient interface, demonstrating the technology's readiness for integration into real-world systems.
\end{enumerate}

An important merit of the proposed approach is its modularity and potential for flexible integration with existing calibration methods. The two stages of the method can be applied not only sequentially but also independently, complementing other techniques. For example, intrinsic camera parameters (focal length, distortion), accurately estimated using classical checkerboard calibration \cite{zhang2000flexible}, can be fixed in the first stage, using the proposed method only for estimating extrinsic parameters. Conversely, the partially defined coordinate correspondence obtained in the first stage can be refined in the second stage not only through interactive manipulation but also using a standard control point alignment procedure if an accurate location plan is available. This flexibility makes the method a valuable addition to the calibration toolkit arsenal, allowing the selection of an optimal hybrid approach for each specific task.

The scientific novelty of the method lies in:
\begin{itemize}
    \item A specially selected set of geometric primitives (parallel, perpendicular, vertical lines, and segments of equal length), sufficient for estimating camera parameters from a single image.
    \item The division of the calibration procedure into two stages: partial parameterization based on geometry and final precise tuning through interactive manipulation, ensuring robustness and accuracy.
    \item Achieving the capability to calibrate systems with minimal overlap of fields of view based on single images, which is a challenging task for many alternative approaches.
\end{itemize}

Thus, the proposed method opens the possibility of deploying accurate multi-camera tracking systems in previously inaccessible or economically non-viable scenarios, significantly expanding the application domain of intelligent video analysis.

\end{document}